\documentclass{article}
\usepackage{tikz}
\usepackage[utf8]{inputenc}
\usepackage[T1]{fontenc}
\usepackage{lmodern}
\usepackage{amsmath}
\usepackage{amssymb}
\usepackage{graphicx}
\graphicspath{{pictures/}}
\usepackage{booktabs}

\usepackage[]{xcolor}
\newcommand{\colorpalette}[1]{%
    \colorlet{#1l}{#1!50!white}%
    \colorlet{#1ll}{#1!20!white}%
    \colorlet{#1lll}{#1!10!white}%
    \colorlet{#1d}{#1!90!black}%
    \colorlet{#1dd}{#1!70!black}%
}
\definecolor{colorf}{HTML}{734f96}%
\colorpalette{colorf}
\definecolor{colorc}{HTML}{1476b5}
\colorpalette{colorc}
\definecolor{colorscorep}{rgb}{0,0.235,0.416}
\colorpalette{colorscorep}
\definecolor{colorfile}{rgb}{0.64,0.44.0}
\colorpalette{colorfile}
\colorlet{highlight}{red!60!black}
\definecolor{goldenhighlight}{HTML}{ffdb57} 
\colorlet{colorno}{red!80!black}
\colorlet{coloryes}{green!80!black}

\usepackage{pifont}

\usepackage{listings}
\usepackage[frozencache,cachedir=.]{minted}
\usepackage{tikz}
\usetikzlibrary{shapes.geometric, arrows, positioning,backgrounds}

\usepackage{acro}
\DeclareAcronym{IPC}{
  short = IPC ,
  long  = instructions per cycle
}
\DeclareAcronym{CI}{
  short = CI ,
  long  = continuous integration
}
\DeclareAcronym{API}{
  short = API ,
  long  = application programming interface
}
\DeclareAcronym{CPT}{
  short = CPT ,
  long  = Critical Path Tool
}
\DeclareAcronym{CG}{
  short = CG ,
  long  = conjugate gradient
}
\DeclareAcronym{MPCDF}{
  short = MPCDF ,
  long  = Max Planck Computing and Data Facility
}
\DeclareAcronym{IO}{
  short = IO ,
  long  = input/output
}
\usepackage{tikz}
\usetikzlibrary{positioning}
\usepackage{pgfplots}

\usepackage[
  a4paper,
  textwidth=117mm,
  textheight=191mm,
  heightrounded,
  hcentering,
  vcentering
]{geometry}

\usepackage{url}
\usepackage{rotating}
\usepackage{placeins}
\usepackage{float}
\newcommand{\todo}[1][]{}

\newcommand{\myabstract}{%
Ensuring good performance is a key aspect in the development of codes that target HPC machines.
As these codes are under active development, the necessity to detect performance degradation early in the development process becomes apparent. In addition, having meaningful insight into application scaling behavior tightly coupled to the development workflow is helpful.
In this paper, we introduce TALP-Pages, an easy-to-integrate framework that enables developers to get fast and in-repository feedback about their code performance using established fundamental performance and scaling factors \cite{wagner_structured_2018}.\\
The framework relies on TALP \cite{lopez_talp_2021}, which enables the on-the-fly collection of these metrics.
Based on a folder structure suited for \ac{CI}, which contains the files generated by TALP, TALP-Pages generates an HTML report with visualizations of the performance factor regression as well as scaling-efficiency tables.\\
We compare TALP-Pages to tracing-based tools in terms of overhead and post-processing requirements and find that TALP-Pages can produce the scaling-efficiency tables faster and under tighter resource constraints.
To showcase the ease of use and effectiveness of this approach, we extend the current \ac{CI} setup of GENE-X \cite{michels_gene-x_2021} with only minimal changes required and showcase the ability to detect and explain a performance improvement.
}

\begin{document}

    \title{TALP-Pages: An easy-to-integrate continuous performance monitoring framework}
    \date{\scriptsize Presented at the 15th International Parallel Tools Workshop 2024\footnote{https://tu-dresden.de/zih/das-department/termine/parallel-tools-workshop-2024}}
    
    \author{Valentin Seitz\textbf{\footnote{Barcelona Supercomputing Center, \texttt{valentin.seitz@bsc.es}} \label{bscref}}, Jordy Trilaksono\footnote{Max-Planck Institute for Plasma Physics, Garching}, Marta Garcia-Gasulla\footnote{Barcelona Supercomputing Center}}

    \maketitle

    \section*{Abstract}
    \myabstract

\section*{Introduction}
\label{sec:talp_pages:introduction}

Developing software for HPC machines is a complex task. 
Apart from the typical challenges during software development, such as ensuring functional correctness, HPC-targeting software needs to make sure to run at acceptable performance and exhibit favorable scaling behavior.  
Writing code that meets this criteria is hard, as scalability bugs or performance degradation might introduce themselves, similar to normal bugs occurring.
To prevent this, fast and easy feedback about potential scaling bugs or performance degradation is essential. \Ac{CI} workflows are already used to provide feedback about functional correctness, e.g. running unit and integration tests. 
Normally, this feedback happens very closely coupled to the development workflow and on the same platform that teams usually collaborate on like GitHub or GitLab.
\tikzstyle{nodebox} = [rectangle, draw, fill=white, text centered, minimum size=1cm, text width=2cm, rounded corners]
\tikzstyle{arrow} = [thick,->,>=stealth]

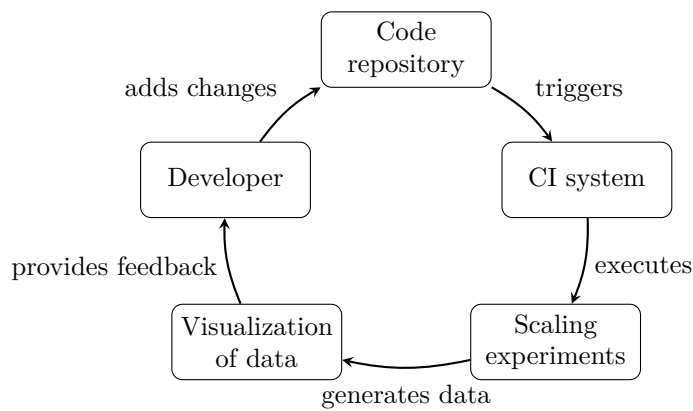
\begin{figure}[h!]
    \centering
    \begin{tikzpicture}[node distance=2.5cm]

    \node[nodebox] (Code) at (90:2.5) {Code repository};
    \node[nodebox] (CI) at (18:2.5) {CI system};
    \node[nodebox] (Performance) at (325:2.4) {Scaling experiments}; 
    \node[nodebox] (Visualization) at (215:2.4) {Visualization of data}; 
    \node[nodebox] (Developer) at (162:2.5) {Developer};

    \draw[arrow] (Code) to[bend left=12] node[midway, above right] {triggers} (CI);
    \draw[arrow] (CI) to[bend left=12] node[midway, below right, xshift=0mm,yshift=2mm] {executes} (Performance);
    \draw[arrow] (Performance) to[bend left=12] node[midway, below, yshift=-1mm] {generates data} (Visualization);
    \draw[arrow] (Visualization) to[bend left=12] node[midway, below left, xshift=0mm,yshift=2mm] {provides feedback} (Developer);
    \draw[arrow] (Developer) to[bend left=12] node[midway, above left] {adds changes} (Code);

    \end{tikzpicture}
    \caption{Abstract workflow for a continuous performance monitoring framework.}
    \label{fig:ci_flow}
\end{figure}

Figure \ref{fig:ci_flow} shows an abstract workflow for continuous performance monitoring using \ac{CI}.
Developers commit code changes into a repository, which triggers a \ac{CI} to run performance experiments generating performance data which then gets visualized and made accessible to the developer. While this approach is nothing new, the critical part of this pipeline is what kind of performance data is gathered and how it is presented to the developers. 
With current approaches mainly focussing on wall clock time as a performance metric, we propose the usage of a well established methodology to get insight into application behavior using fundamental performance factors \cite{wagner_structured_2018}.

In this paper, we introduce a framework that easily allows developers to track performance regression of their code and get insight into the code's performance using these fundamental performance factors.
All this happens closely coupled to their development workflow in the same repository and without additional hardware resources required.
This is enabled by calculating the performance model-factors on the fly, saving them as \verb|json| files and visualizing them later in an interactive report.

With TALP-Pages, we contribute to the currently available \ac{CI} frameworks a tool that:
\begin{itemize}
    \item does not require additional hardware only some CI integration that has access to the target machine
    \item provides insightful metrics that enable identification of scaling problems
    \item can be applied to applications without requiring code or compilation changes
    \item provides a tracing-free way to obtain a full table with scaling-efficiency metrics
\end{itemize}

First, we discuss current work in \ac{CI} based performance monitoring, followed by an introduction of the fundamental performance factors and the performance tools able to obtain them.
The next section introduces the framework and explains the different sub-components and provides a details on how to use it in an \ac{CI} setting.
We then compare the the framework to other tools able to obtain similar performance metrics followed by an exemplary integration into a plasma physics application.

\section*{Background}
\subsubsection*{CI Frameworks}
Recently, a lot of teams developed tailor made solutions to enable a continuous performance monitoring catering to the special need of their particular software design workflow and reflecting their organizational capabilities.
In general, the problem of assessing performance continuously can be broken down in the following aspects: What performance data is collected, how this performance data is stored and how data is presented to the developers.

A common solution is to measure the wall clock time for a certain problem and track the evolution of that. With this approach, identifying the cause of a regression might be difficult, as no additional information other than wall-clock time is available. 
For more refined timing information, internal profiling tools that measure the timings for certain code regions can be employed. This aids identifying the region responsible for a performance regression.

Anzt et al. \cite{anzt_towards_2019} collects the wall-clock timings and memory usage of certain problems and stores them in a secondary repository organized by architecture and benchmarking case. The code is executed only after a thorough review to mitigate potential security issues through unknown code execution. The resulting files contain not only the wall-clock time, but also input-parameters aiding better reproducibility. 
The results are visualized through a custom web-app hosted trough GitHub. 

Instead of storing the performance results in a repository, \cite{alt_continuous_2024} developed a continuous-benchmarking architecture that relies on an time-series database to store the performance data. This data is later visualized using open-source software which is hosted on a dedicated web-server.
Here, the performance data not only contains the time-to-solution, but also some custom metrics meaningful to the type of code this pipeline is used for. 
Additionally, some hardware counters are read using LIKWID \cite{treibig_likwid_2010}.
The framework also collects information about the environment where the execution happened, to enhance reproducibility.

Unspecified performance data was gathered by \cite{dosimont_monitoring_2024}, which presented a closed-source framework that has highlighted some platform instabilities and misconfigurations in the HPC system it was run on. The data is stored similar to \cite{alt_continuous_2024} in a database.

To summarize, most of the above presented solutions require additional infrastructure like databases or web servers to provide the storage and visualization of the gathered performance data.
Some frameworks store additional hardware counters or metadata about the machine and input case for reproducibility. Most frameworks provide only regression information about wall-clock time for certain benchmarking cases.

While wall-clock time evolution is a good fit to track regression, it lacks the ability to explain a performance change.

Here, fundamental performance factors visualized using scaling-efficiency tables can help developers to identify problems and explain performance changes.

\subsubsection*{Scaling-efficiency Table}
The scaling-efficiency table consists of the fundamental performance factors which are gathered for each resource configuration the application was executed in.
These factors that are introduced in \cite{wagner_structured_2018} can be split in two parts, which make up the the global efficiency of the application.
Computation scalability tracks how the amount of work (the sum of executed instructions) and how the rate at which this work is processed (frequency and \ac{IPC}) are changing across the different executions. This metric is defined with respect to a reference case.
Parallel efficiency on the other hand is an absolute measure quantifying how good the parallelization, e.g. the usage of the programming models, of the application is.
This metric can be further split into a hierarchy that attributes efficiency losses to different programming models and their particularities. Load balance, serialization, or losses by communication times are a typical example for these factors.
As these numbers represent efficiencies, they normally range between zero and one, which provides an intuitive approach to understand complex parallel application behavior.

Current tracing-based tools \cite{brunst_score-p_2012, geimer_scalasca_2010} \cite{noauthor_extrae_nodate,pillet_paraver_nodate} that enable the computation of these fundamental performance factors impose high requirements in terms of memory, storage and time that render this option infeasible to be used in a \ac{CI} environment.
Tools like the \ac{CPT} \cite{schwitanski_--fly_2022} which provides the fundamental performance factors on the flight, cannot provide the full table as it does not read hardware counters, which disables insight on the computation scalability of the application. 

The TALP \cite{lopez_talp_2021} module of DLB is the only tool that currently provides the metrics including hardware counters using on-the-fly calculation.

\subsection*{TALP module}
TALP currently supports the calculation of MPI metrics and introduced OpenMP metrics in version 3.5.0 as an experimental feature.
By default, TALP creates a \verb|Global| region capturing the whole program execution. TALP also provides an \ac{API} that allows for annotation of regions inside the code. This is used to compute the performance metrics on a per-region basis, allowing more fine-grained insight into the application. 
The currently supported applications are MPI and OpenMP applications written in C/C++ or Fortran.

\section*{TALP-Pages}
TALP-Pages can either be used as a standalone tool on the user's machine or integrated into a \ac{CI} workflow.
The usage is similar, so we will first introduce the standalone usage and then continue to highlight the proposed usage in a \ac{CI} system.

As a first step, the user selects the resource configuration, namely the combination of MPI processes and OpenMP threads together with suitable input cases that are of interest. These performance runs normally consist of weak or strong scaling experiments but can also encompass comparisons between different resource configurations for a given input.

Aftr this, the user runs the application with DLB TALP, which generates a \verb|json| file containing the performance data. Note, that the application does not need to be modified or compiled with instrumentation.
After obtaining the \verb|json| files for the different performance runs, the user organizes them semantically with the following constraints:
\begin{itemize}
    \item All \verb|json| files in a folder should be part of a weak or strong scaling experiment or be a comparison between resource configurations
    \item There needs to be a top-level folder containing the experiment-specific folders.
    \item Previous runs of the same experiment must go in the same folder.
\end{itemize}
A valid folder structure running two different input meshes can be seen in listing \ref{code:tree_of_things}.

\begin{figure}
\begin{small}
\begin{verbatim}
`-- talp_folder
    |-- mesh_1
    |   |-- comparison
    |   |   |-- talp_1x112.json
    |   |   |-- talp_2x56.json
    |   |   `-- talp_4x28.json
    |   |-- strong_scaling
    |   |   |-- talp_8x14.json
    |   |   `-- talp_8x28.json
    `-- mesh_2
        |-- weak_scaling
            |-- talp_8x14_9dc04ca.json
            |-- talp_8x28_9dc04ca.json
            |-- talp_8x14_ed8b9ef.json
            `-- talp_8x28_ed8b9ef.json

\end{verbatim}
\end{small}
\caption{Exemplary folder structure suitable as TALP-Pages input.}
\label{code:tree_of_things}
\end{figure}

After placing the \verb|json| files into the correct folder hierarchy, the user can generate an HTML report by pointing TALP-Pages to the top-level folder containing the different experiments like: \verb|talp ci-report -i ./talp_folder -o output|. 

The generated HTML report consists of three main components: Time-evolution plots (\ref{subsec:time_series_plots}) for each resource configuration, a scaling-efficiency table (\ref{subsec:scaling_effiency_table}) for each sub-folder, and a SVG badge displaying the parallel efficiency for each resource configuration.
If the TALP \ac{API} was used, the user can specify the region for which scaling-efficiency tables should be generated. The selected regions are also highlighted in the time-series plots.

\subsection*{Time-evolution plots}
\label{subsec:time_series_plots}
The time-series plots are generated for every resource configuration found in a sub-folder.
Since time-series plots require a timestamp the timestamp provided by DLB, which corresponds to the end time of the execution is used.
If additional git metadata is present in the \verb|json|, the git commit timestamp is used instead.
The time-series plots consists of a plot for the elapsed time per-region, another row for the computation related performance indicators like \ac{IPC}, frequency and number of instructions, and additional rows for the evolution the parallel effiency and the availabel sub-metrics.
To reduce optical cluttering, the regions can be easily toggled on and off interactively on the generated report, such that individual regions of interest can be easily tracked.
An example of these time-series plots can be seen in Figure \ref{fig:time_series_report}.

\subsection*{Scaling-efficiency table}
\label{subsec:scaling_effiency_table}

\begin{figure}
    \centering
    \includegraphics[width=1\linewidth]{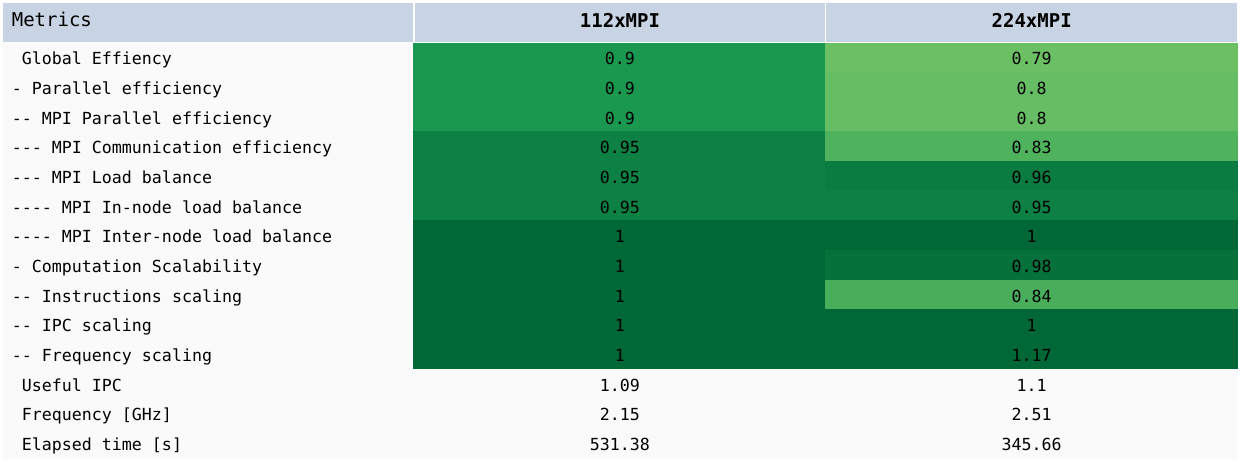}
    \caption{Scaling efficiency table generated by TALP-Pages of a MPI-only strong scaling experiment}
    \label{fig:scaling_effiency_table}
\end{figure}
The scaling-efficiency table shown in Figure \ref{fig:scaling_effiency_table} is generated once per sub-folder.
For each resource configuration, the latest timestamp is taken and a column containing the parallel efficiency metrics, as well as the computational scaling metrics are generated.
As the computational scaling metrics like \ac{IPC} scaling, instruction scaling and frequency scaling need to be computed relative to a reference, we select the resource configuration with the least resources to be that reference case. 
For the detection of the scaling mode, we assume that for weak scaling the instructions executed per CPU are constant. If this condition is violated, we detect strong scaling.
The scaling mode only influences the computation of the instruction scaling.
For strong scaling experiments, we assume the total number of instructions reported by TALP to stay constant and any deviation therefrom will be seen as inefficiency, whereas for weak scaling this assumption is made for the instructions per CPU.

Thus far, we showed the general usage of TALP-Pages and its visualization outputs. 
Let us now consider a possible \ac{CI} workflow using TALP-Pages.

\subsection*{CI Workflow}
As TALP-Pages is designed with \ac{CI} in mind, the folder structure introduced in \ref{code:tree_of_things} is very suitable for modern CI/CD solutions and the respective artifact management they provide.
Compared to other frameworks, that require an externally hosted database or an additional repository, TALP-Pages only relies on the artifacts management capabilities of \ac{CI} systems to store and retrieve past performance data.

\begin{figure}
    \centering
    \includegraphics[width=1\linewidth]{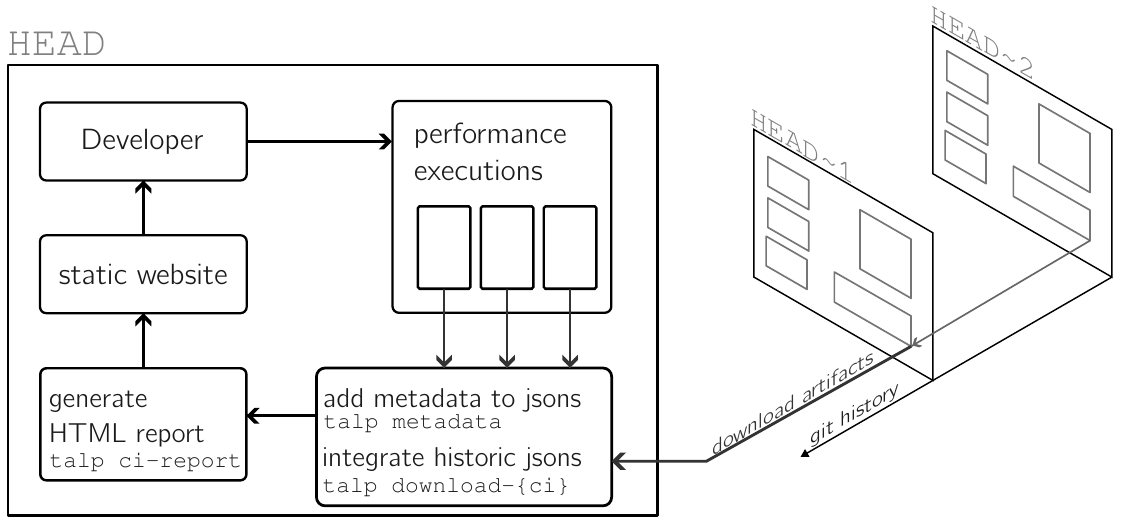}
    \caption{Exemplary usage of TALP-Pages in a \ac{CI} environment}
    \label{fig:talp_pages_ci_flow}
\end{figure}

Looking at \ref{fig:talp_pages_ci_flow}, we start with the developer who triggers the \ac{CI} to run the pre-defined performance jobs. Here, a runner with access to a dedicated machine is needed.
After that, we provide a convenience wrapper to automatically add git-related metadata like commit hash, branch name and timestamp of the commit to the \verb|json| generated by TALP.
We collect the newly generated \verb|json| files via the native artifacts handling in an accumulating job and additionally download the previous pipelines artifacts and copy them over into the current structure.
For the downloading of the artifacts, we also provide a small wrappers for common CI solutions, to hide these details from the users.

After the \verb|json| files are in the folder structure, which now contains the results of historic and current runs, we can invoke TALP-Pages and generate the interactive report.

This HTML report can be hosted in an in-repository static website hosting solution. This makes it easily discoverable by the developer and does not require additional hardware.

For the next cycle of the \ac{CI}, the \verb|json| files added in this cycle will be contained in the artifacts downloaded. 
With this technique we exchanged the previously required external database or separate repository with the artifacts management of the \ac{CI} solution in use. 

Having introduced TALP-Pages and its capabilities, we will now first compare it to other performance analysis tools in terms of overheads during runtime of the application and resource requirements to generate the scaling-efficiency table. 

\section*{Comparison to other tools}
\label{sec:talp_pages_comparison}
As layed out in \cite{wagner_structured_2018}, obtaining the scaling-efficiency table is an important first step to identify fundamental scaling problems of the application.
Currently there are four, including TALP-Pages, tool-suites known to the authors that can compute at least the partial table.
In the following we will first quickly introduce the tools and describe their usage, followed by an analysis of the overheads in terms of memory, storage and processing time needed to obtain the scaling-efficiency table.

The Score-P/Scalasca \cite{brunst_score-p_2012,geimer_scalasca_2010} also known as the JSC-toolset, provides the fundamental performance factors by post-processing an application trace and profiling runs generated by Score-P.
The trace gets post-processed by Scalasca. Later the tracing results and profiling results get merged into a file that can be explored by Cube \cite{saviankou_cubegui_2023}. From Cube the fundamental performance factors can be extracted and brought into a tabular format, which involves some manual work. When enabling instrumentation in Score-P, these factors can be easily obtained for the different regions.

Similarly, the BSC-Toolset consisting of Dimemas, Extrae,  Paraver and Basicanalysis \cite{noauthor_extrae_nodate, pillet_paraver_nodate} performs a post-mortem analysis of an application trace generated by Extrae.
Here the whole table is automatically generated in multiple formats by Basicanalysis. To get the performance factors for different regions, the application trace needs to be cut prior to invoking the basic analysis tool to only contain that region, which involves manual work.

In contrast to that, the \ac{CPT} \cite{schwitanski_--fly_2022} provides the fundamental performance factors by an online computation directly after the execution of the application ended.

As the \ac{CPT} currently lacks Fortran support, we chose a C++ based MPI+OpenMP mini-app \cite{martineau_assessing_2017} to evaluate the different tools.
The exact environments and versions used to build the application and the tools in MareNostrum 5 can be found in the appendix in sections \ref{sec:software_env} and \ref{sec:tealeaf_benchmark}.
We ran a strong application scaling experiment of the \ac{CG} solver with a 2D grid resolution of $4000^2$ points for four time steps.
We scaled from one node in MareNostrum 5 to two nodes using 1 MPI process per socket and 56 OpenMP threads, with each of the OpenMP threads pinned to a core on the same socket as the corresponding MPI processes was placed. Simultaneous multithreading was disabled for all experiments.

Additionally, we run a weak scaling experiment with the same \ac{CG} solver starting from one node to an increased resolution of $8000^2$. As this results in four times the amount of cells, we use four nodes for the second run.
For the two experiments, we reused the runs from the initial case executing in one node.

As our goal is to obtain the scaling-efficiency table, we now compare TALP-Pages to other tools. To do this, we first classify the runtime overhead in terms of time followed by an overview of resource requirements to generate the scaling-efficiency table for the whole execution.
The generated tables can be found in section \ref{sec:scaling_effiency_tables}.

\subsection*{Runtime Overhead}
The tracing libraries Score-P and Extrae were configured to only trace the bare minimum required to obtain the fundamental performance factors.

\begin{table}[!t]
\renewcommand{\arraystretch}{1.2}
\begin{tabular}{p{1.3cm}p{1.3cm}p{2.0cm}p{1.0cm}p{1.0cm}p{1.2cm}p{1.0cm}}
Problem size & Resources {\tiny (MPIxOpenMP)}  & runtime {(stddev)} & \multicolumn{4}{c}{runtime overhead}  \\
\cline{4-7}
\noalign{\smallskip}
& & & DLB & \ac{CPT} & Score-P & Extrae \\
\noalign{\smallskip}\hline\noalign{\smallskip}
$4000^2$ & 2x56 & 125s (0.1\%) & 4.7\% & 2.5\% & 2.4\% & 5.4\% \\
$4000^2$ & 4x56 & 27s (0.5\%) & 22\% & 14\% & 11\% & 23\% \\
$8000^2$ & 8x56 & 266s (0.2\%) & 5.9\% & 4.1\% & 3.3\% & 7.8\% \\
\noalign{\smallskip}\hline\noalign{\smallskip}
\end{tabular}

\caption{Average measured runtime overheads of the different tools executing }
\label{tab:runtime_overhead}  
\end{table}

Looking at table \ref{tab:runtime_overhead} we can see that the \ac{CPT} and Score-P have the lowest runtime overhead. It should be mentioned, that for the \verb|POP| preset used in this analysis the Score-P runs the application twice, which might help them maintain a low overhead, as in only one of the executions hardware counters are collected. 
Similarly the \ac{CPT} does not collect hardware counters, which might be an indicator why its performs similarly to Score-P and in general better than tools that collect hardware counters.

In the second row, we can see a very high overhead for all tools. This can be attributed to the fine granularity, especially in OpenMP constructs found when scaling the application strongly. In general, these results should be omitted and a more suitable target should be chosen, as high overheads can lead to significantly skewed results. But as our goal is not the application assessment, but rather the tools, we keep the data-point to provide a "worst-case" from the tools perspective.

In order to obtain the table containing the fundamental performance factors, we will now follow each tool-specific process and asses the imposed post-processing requirements

\subsection*{Post-processing requirements}
We classify the post-processing needs in 3 resource categories. One is memory that the tool needs to have in order to generate the table, another one is storage and the third processing time. If the creation of the table involves multiple steps, we take the maximum of resources consumed during one of the steps, as this determines the minimum resource requirements. We don't classify minimal manual labor involved e.g. combining results provided in a table.
The post-processing steps of the JSC-tools were run with the same resources available as the initial execution.
All other post-processing executions were done in a single MareNostrum 5 node.

\begin{table}[!t]
\renewcommand{\arraystretch}{1.2}
\setlength{\tabcolsep}{8pt} 
\begin{tabular}{p{2cm}|ll|ll|ll}
 & \multicolumn{2}{l|}{Memory [GB]} & \multicolumn{2}{l|}{Storage [GB]} & \multicolumn{2}{l}{Time [s]} \\
\hspace{0.90cm}Scaling & weak & strong & weak & strong & weak & strong \\ 
\hline
TALP-Pages & 0.13 & 0.13 & 0.02 & 0.02 & 2 & 2 \\
JSC-Tools & 44 & 19 & 29 & 6.7 & 436 & 441 \\
BSC-Tools & 138 & 32 & 165 & 49 & 10.8 $\times10^3$ & 3.03 $\times10^3$
\end{tabular}
\caption{Minimum resource requirements and processing time for different tools generating the scaling-efficiency table}
\label{tab:minimum_requirements} 
\end{table}

Looking at table \ref{tab:minimum_requirements} we can see that we excluded the \ac{CPT} for this analysis, as it mainly involved only minimal manual labor to obtain the scaling table by copying together the resulting files.

The significant difference in processing time of the BSC-Tools compared to the JSC-tools can be mainly explained by the time consumed by Dimemas, a sequential tool that uses simulations of MPI communications to split the MPI Communication efficiency even further.
Table \ref{tab:minimum_requirements} provides evidence, that the trace-based tool sets from BSC and JSC require orders of magnitude more resources to generate the scaling-efficiency table.

These findings suggest that compared to the trace based tools, TALP-Pages provides an acceptable run-time overhead, but significantly smaller post-processing requirements. This makes it an attractive solution for users, who are only interested in obtaining the tables, without any following detailed trace-based analysis. This becomes especially important in resource constraint environments like \ac{CI}.

As the four tools presented here have very different goals it may seem off to portray them together. In our comparison, we set the goal to generate the scaling-efficiency table. Doing this very efficiently was most likely not a design goal of tracing tool suites like the JSC and BSC one.

After comparing TALP-Pages with other tools obtaining the fundamental performance factors, we now showcase the integration of TALP-Pages into an application.

\section*{Integration into GENE-X}
\subsection*{GENE-X}
GENE-X \cite{michels_gene-x_2021} is a plasma turbulence code developed mainly at the Max Planck Institute for Plasma Physics (IPP). Its main use case is to simulate the turbulent plasma behavior in any region from the core of the plasma to the walls of the magnetic confinement fusion device. 

GENE-X is written in modern Fortran 2008. It uses MPI and OpenMP for Hybrid CPU parallelism.
The GENE-X team has a particular focus on good software engineering practices and already relies on CI to do integration and unit testing for the software.

In order to integrate TALP-Pages with GENE-X, some changes to the code were necessary.
GENE-X already had a built-in profiling tool based on nested regions \cite{michels_gene-x_2021}. This made the integration of the TALP \ac{API} straightforward. 
Additionally, we modified the CMake setup of GENE-X to allow optional building of the TALP annotations.

\subsection*{GitLab Setup}
As far as the \ac{CI} component is concerned, GENE-X uses a GitLab instance hosted at the \ac{MPCDF} with GitLab Pages capabilities.
We registered private GitLab runners on MareNostrum 5 and Raven -a cluster also hosted by \ac{MPCDF}- to run the performance jobs. 

To integrate TALP-Pages we added a new stage to the pipeline which executes the performance measurements called \verb|performance|.
Secondly, we introduced the \verb|talp-pages| job that accumulates the performance data and generates the HTML report. 

\begin{figure}
\begin{minted}[linenos=true,fontsize=\small]{yaml}
.performance-template: &performance_job
    stage: performance
    tags:
        - $MACHINE_TAG
    script:
        - *load_genex_modules
        - *build_genex
        - sbatch --wait ./${CONFIGURATION}/submit.sh
        - cp talp.json talp/${CASE}/${RESOLUTION}/${MACHINE_TAG}/...
        talp_$(date+"%FT%H%M")_${CI_COMMIT_SHORT_SHA}.json
    artifacts:
      paths:
        - talp

performance-cpu-fast:
  <<: *performance_job
  variables:
    WITH_DLB: "true"
    CASE: "salpha"
    ...
  parallel:
    matrix:
      - RESOLUTION: "resolution_2"   
        CONFIGURATION: ["1Nx2MPI","2Nx4MPI"]
        MACHINE_TAG: ["raven","mn5"]
\end{minted}
\caption{Matrix job example of the the performance stage in the GENE-X pipeline}
\label{code:talp_cpu_fast_job}

\end{figure}
In the performance stage, we rely on yaml anchors and variables to allow for easy instantiation of multiple jobs. Looking at listing \ref{code:talp_cpu_fast_job}, we can see how the \verb|performance_job| anchor gets re-used in a matrix job called \verb|performance-cpu-fast|. In line 9 of \ref{code:talp_cpu_fast_job} the generated \verb|talp.json| is copied directly into the correct folder structure. Later (line 11-13) the whole \verb|talp| folder is exported as an artifact.
For the \verb|performance-cpu-fast| job, we run a strong scaling experiment of the \verb|salpha| input in resolution \verb|resolution_2| from one node to two nodes in MareNostrum 5 and Raven.

Additional to the performance stage, we introduced the \verb|talp-pages| job.
As can be seen in listing \ref{code:talp_pages_job}, the artifacts are accumulated in this job and the artifacts of the previous execution are downloaded.
The downloaded artifacts are unzipped and copied over. 
After this, we generate the HTML report, selecting the \verb|timestep| and \verb|initialize| regions which were instrumented with the TALP \ac{API}.

\begin{figure}
\begin{minted}[fontsize=\small]{yaml}
talp-pages:
  stage: deploy
  needs:
    - job: performance-cpu-fast
      artifacts: true
  script:
    - talp metadata -i talp # Add metadata to newly generated jsons
    - talp download-gitlab ... --output-file talp_history.zip
    - unzip talp_history.zip # Unzip downloaded artifacts
    - cp -r talp_history/* talp # Copy over history
    - talp ci-report -i talp -o public/talp 
      --regions initialize timestep --region-for-badge timestep
  artifacts:
    paths:
      - public/talp # GitLab Pages
      - talp # To download the next pipeline execution

\end{minted}
\caption{Shortened talp-pages job definitions used in the GENE-X pipeline}
\label{code:talp_pages_job}
\end{figure}
\subsection*{Reports}

To check if TALP-Pages is able to detect changes in performance, we fixed a scaling bug discovered in a previous assessment of the code.
\begin{figure}
    \centering
    \includegraphics[width=1\linewidth]{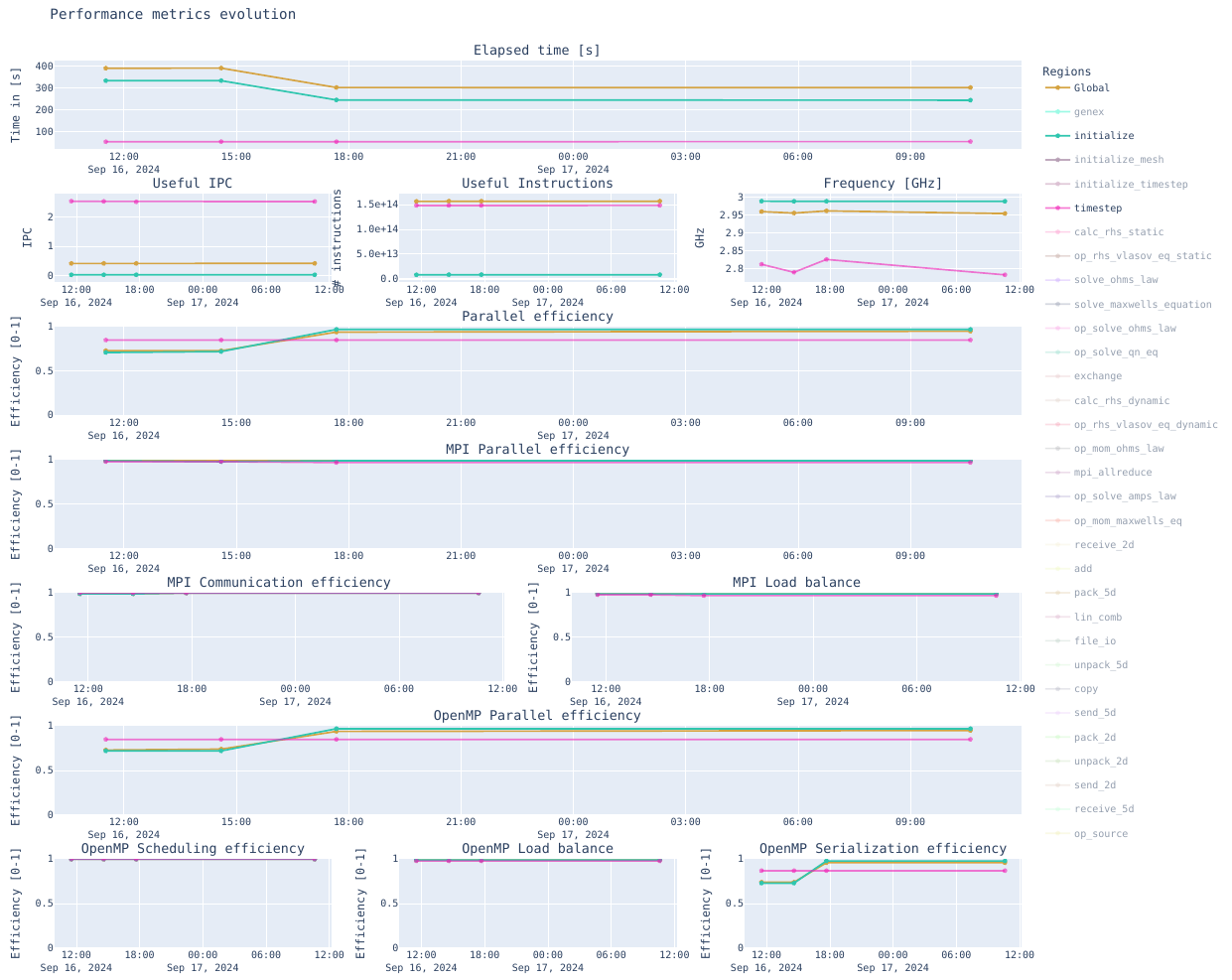}
    \caption{Time series plot generated in the GENE-X pipeline running the salpha case in resolution\_3. Results obtained in MareNostrum 5 with 8 MPI processes and 56 OpenMP threads.}
    \label{fig:time_series_report}
\end{figure}

Looking at figure \ref{fig:time_series_report} we can see the runtime improvement with elapsed time decreasing for the \verb|initialize| and the implicit region \verb|Global|.
The \verb|timestep| region seems unaffected. 
Looking at the computational metrics in the second row of the figure, we can see that neither \ac{IPC}, nor instruction or frequency changed considerably from the previous executions to warrant the drop in execution time. 
Actually, the parallel efficiency of the application as reported by TALP increased significantly, leading to an overall speedup of the \verb|inialize| region. 
Checking the child metrics by going down the rows in the figure, we can see that OpenMP serialization efficiency is responsible for the parallel efficiency increase.

Additional to the time-series plots, we also obtained the scaling-efficiency tables for the strong-scaling experiments for both Raven and MareNostrum 5.

Even though TALP-Pages has been partially integrated into the CI of GENE-X, these changes are still pending to be merged into the main branch and become productive.

\section*{Discussion}
A big problem for the wide-spread adoption of continuous performance monitoring is the limited organizational support by HPC operators, as already pointed out by \cite{gamblin_overcoming_2022}.
TALP-Pages relies on runners being able to execute code on HPC-machines. Having these runners set up on the login nodes as we did for this paper might work for some teams, but permission and security concerns quickly arise. A wider spread adoption of solutions like Jacamar \cite{bryant_jacamar_nodate} might help to overcome this issue.

As we rely on TALP for gathering the performance factors, which internally relies on tool interfaces like \verb|OMPT|, we currently don't support the OpenMP metrics for the compilers like \verb|gcc|, as not all runtimes implement the \verb|OMPT| interface. While there are workarounds for some compilers, these might not be very obvious to application developers.
Another deficiency of TALP currently is the blindness to IO operations. When running without instrumentation, the \ac{IO} part of the application might introduce high variance between the runs which skews collected performance factors significantly. 
We therefore recommend to instrument \ac{IO} intensive regions with the TALP \ac{API} to get a clearer view if \ac{IO} performance fluctuates between runs.
Another problem, that might arise is that with enough data the artifacts management of the \ac{CI} solution could become inadequate. Here alternative storage solutions like cloud storage could offer a solution.

\section*{Conclusions and future work}
In this paper, we presented a framework that provides insight into fundamental performance factors of applications within a \ac{CI} environment. This is achieved by using TALP to gather the fundamental performance factors during runtime. Then store the resulting \verb|json| files and employ the artifact management of the CI system to retrieve them at a later point for regression insight. Additional to the performance regression, TALP-Pages offers insight into the scaling behavior of the application using well established scaling-efficiency tables.
When compared to other tools that generate the scaling-efficiency table, TALP-Pages is able to compute these tables under tighter resource constraints.
We showcased the integration of TALP-Pages into GENE-X with only little code changes and demonstrated that TALP-Pages is able to detect performance improvements of the code in an explainable way.

With this in mind, the next steps would be to provide an exemplary integration of TALP-Pages using GitHub. As soon as DLB TALP provides support for GPUs, we plan to integrate that as well. 
Additionally, exporting an Extra-P \cite{calotoiu_using_2013} experiment from a collection of \verb|jsons| could be an interesting addition to be able to extend the performance modeling capabilities.

\newpage
\section*{Acknowledgement}
Special thanks go to Thomas Gruber and the conversation at the sidelines of the deRSE 2024 and access to his GitLab prototype.\\
This work has been supported by the European High-Performance Computing Joint Undertaking (JU) under grant agreement No 101093038 (ChEESE-2P).

\section*{Appendix}

\subsection*{Software environment}
\label{sec:software_env}
\FloatBarrier
\begin{table}[ht]
\centering
\renewcommand{\arraystretch}{1.2}
\setlength{\tabcolsep}{8pt} 
\begin{tabular}{l|l|l}
Software & Version & Module in MareNostrum 5\\ 
\hline
Intel OneAPI Compilers & 2023.2.0 & intel/2023.2.0 \\
Intel MPI & 2021.10.0 & impi/2021.10.0 \\
UCX & 1.16.0 & ucx/1.16.0 \\
PAPI & 7.1.0 & papi/7.1.0-gcc
\end{tabular}
\caption{Module environment used to build the tools and benchmarking application in MareNostrum 5}
\label{tab:module_enviroment}
\end{table}
\FloatBarrier

\begin{table}
\centering
\renewcommand{\arraystretch}{1.2}
\setlength{\tabcolsep}{8pt} 
\begin{tabular}{lll}
Software & Version & URL to source  \\ 
\hline
Score-P & 8.4 & https://doi.org/10.5281/zenodo.10822140  \\
Scalasca & 2.6.1 & https://doi.org/10.5281/zenodo.7440854  \\
CubeGUI & 4.8.2 & https://doi.org/10.5281/zenodo.8345207 \\
Extrae & 4.2.3 +1 & https://github.com/valentin-seitz/extrae/tree/fix-segfault-omp
\end{tabular}
\caption{Manually installed software in MareNostrum 5.}
\label{tab:additional_software}
\end{table}

\begin{table}
\centering
\renewcommand{\arraystretch}{1.2}
\setlength{\tabcolsep}{8pt} 
\begin{tabular}{llll}
Software & Version & URL to source \\ 
\hline
Paraver & 4.11.4 & https://ftp.tools.bsc.es/wxparaver/wxparaver-4.11.4-src.tar.bz2 \\
Dimemas & 5.4.2-devel & https://github.com/bsc-performance-tools/dimemas/tree/317a2d0 \\
Basicanalysis & 0.3.9 & https://ftp.tools.bsc.es/basicanalysis/basicanalysis-0.3.9-src.tar.bz2\\
DLB TALP & 3.5.0-rc1 & https://github.com/bsc-pm/dlb/releases/tag/v3.5.0-rc1 
\end{tabular}
\caption{Software provided in MareNostrum 5 by modules during the experiments}
\label{tab:module_software}
\end{table}

\subsection*{TeaLeaf Benchmark}
\label{sec:tealeaf_benchmark}
We used the TeaLeaf benchmark accessible under: \url{https://github.com/UoB-HPC/TeaLeaf} at commit \verb|e70261c| from the 4.12.2023.

For the measurements with Extrae, CPT and TALP we used the same binary produced by configuring the project with \verb|-DMODEL=omp -DENABLE_MPI=ON|. 

For the Score-P binary, we used the Score-P compiler wrappers with the following arguments: \verb|--nomemory --nouser --nocompile|.
The \verb|CMAKE_BUILD_TYPE| was \verb|release| for the two binaries.

\subsection*{Scaling-efficiency tables}
\label{sec:scaling_effiency_tables}
\FloatBarrier

\begin{sidewaystable}
\centering
\begin{tabular}{l|ll|ll|ll|ll}
 & \multicolumn{2}{c|}{BSC } & \multicolumn{2}{c|}{CPT:} & \multicolumn{2}{c|}{JSC} & \multicolumn{2}{c}{TALP-Pages} \\
  \hspace{2.5cm}Ressources ({\small MPIxOpenMP})& 2x56 & 8x56 & 2x56 & 8x56 & 2x56 & 8x56 & 2x56 & 8x56 \\ 
\hline
Global efficiency & 0.90 & 0.42 & - & - & 0.89 & 0.42 & 0.91 & 0.42 \\
- Parallel efficiency & 0.90 & 0.86 & 0.90 & 0.85 & 0.89 & 0.85 & 0.91 & 0.87 \\
-- Load balance & 0.97 & 0.94 & 0.96 & 0.93 & 0.96 & 0.95 & - & - \\
-- Communication efficiency & 0.94 & 0.92 & 0.94 & 0.91 & 0.93 & 0.90 & - & - \\
- Computation scalability & 1.00 & 0.48 & - & - & 1.00 & 0.49 & 1.00 & 0.49 \\
-- IPC scalability & 1.00 & 1.00 & - & - & 1.00 & 1.02 & 1.00 & 1.00 \\
-- Instruction scalability & 1.00 & 0.49 & - & - & 1.00 & 0.48 & 1.00 & 0.49 \\
-- Frequency scalability & 1.00 & 0.99 & - & - & 1.00 & 0.99 & 1.00 & 0.99 \\
-- Hybrid Parallel efficiency & 0.90 & 0.86 & 0.90 & 0.85 & 0.89 & 0.85 & 0.91 & 0.87 \\
--- MPI Parallel efficiency & 0.98 & 0.94 & 0.97 & 0.93 & 0.96 & 0.93 & 1.00 & 0.99 \\
---- MPI Load balance & 1.00 & 0.98 & 1.00 & 0.98 & 0.99 & 0.98 & 1.00 & 0.99 \\
---- MPI Communication efficiency & 0.98 & 0.96 & 0.97 & 0.95 & 0.97 & 0.95 & 1.00 & 1.00 \\
----- MPI Serialization efficiency & 0.99 & 0.98 & 0.99 & 0.99 & - & - & - & - \\
----- MPI Transfer efficiency & 0.99 & 0.98 & 0.98 & 0.96 & - & - & - & - \\
--- OpenMP Parallel efficiency & 0.93 & 0.91 & 0.93 & 0.92 & 0.93 & 0.92 & 0.91 & 0.87 \\
---- OpenMP Load Balance & 0.97 & 0.95 & 0.96 & 0.96 & 0.97 & 0.97 & 0.99 & 0.98 \\
---- OpenMP Communication efficiency & 0.96 & 0.96 & 0.97 & 0.97 & 0.96 & 0.95 & - & - \\
----- OpenMP Serialisation Efficiency & - & - & 0.99 & 0.99 & - & - & - & - \\
----- OpenMP Transfer Efficiency & - & - & 0.98 & 0.98 & - & - & - & - \\
---- OpenMP Scheduling efficiency (TALP only) & - & - & - & - & - & - & 0.99 & 0.99 \\
---- OpenMP Serialization efficiency (TALP only) & - & - & - & - & - & - & 0.94 & 0.90
\end{tabular}
\caption{Weak scaling-efficiency tables obtained by the different tools.}
\end{sidewaystable}

\begin{sidewaystable}
\centering
\begin{tabular}{l|ll|ll|ll|ll}
 & \multicolumn{2}{c|}{BSC-Tools} & \multicolumn{2}{c|}{CPT} & \multicolumn{2}{c|}{JSC-Tools} & \multicolumn{2}{c}{TALP-Pages} \\
 \hspace{2.5cm}Ressources ({\small MPIxOpenMP})& 2x56 & 4x56 & 2x56 & 4x56 & 2x56 & 4x56 & 2x56 & 4x56 \\ 
\hline
Global efficiency & 0.90 & 1.70 & - & - & 0.89 & 1.92 & 0.91 & 1.8 \\
- Parallel efficiency & 0.90 & 0.63 & 0.90 & 0.58 & 0.89 & 0.58 & 0.91 & 0.63 \\
-- Load balance & 0.97 & 0.82 & 0.96 & 0.85 & 0.96 & 0.86 & - & - \\
-- Communication efficiency & 0.94 & 0.76 & 0.94 & 0.68 & 0.93 & 0.67 & - & - \\
- Computation scalability & 1.00 & 2.71 & - & - & 1.00 & 3.32 & 1 & 2.85 \\
-- IPC scalability & 1.00 & 3.10 & - & - & 1.00 & 3.67 & 1 & 3.28 \\
-- Instruction scalability & 1.00 & 0.98 & - & - & 1.00 & 1.03 & 1 & 0.99 \\
-- Frequency scalability & 1.00 & 0.89 & - & - & 1.00 & 0.88 & 1 & 0.88 \\
-- Hybrid Parallel efficiency & 0.90 & 0.63 & 0.90 & 0.58 & 0.89 & 0.58 & 0.91 & 0.63 \\
--- MPI Parallel efficiency & 0.98 & 0.89 & 0.97 & 0.82 & 0.96 & 0.82 & 1 & 0.96 \\
---- MPI Load balance & 1.00 & 0.97 & 1.00 & 0.97 & 0.99 & 0.95 & 1 & 0.96 \\
---- MPI Communication efficiency & 0.98 & 0.92 & 0.97 & 0.85 & 0.97 & 0.87 & 1 & 1 \\
----- MPI Serialization efficiency & 0.99 & 0.98 & 0.99 & 0.96 & -  & - & - & - \\
----- MPI Transfer efficiency & 0.99 & 0.94 & 0.98 & 0.88 & -  & - & - & - \\
--- OpenMP Parallel efficiency & 0.93 & 0.70 & 0.93 & 0.70 & 0.93 & 0.70 & 0.91 & 0.63 \\
---- OpenMP Load Balance & 0.97 & 0.84 & 0.96 & 0.87 & 0.97 & 0.91 & 0.99 & 0.96 \\
---- OpenMP Communication efficiency & 0.96 & 0.83 & 0.97 & 0.81 & 0.96 & 0.77 & - & - \\
----- OpenMP Serialisation Efficiency & - & - & 0.99 & 0.93 & - & - & - & - \\
----- OpenMP Transfer Efficiency & - & - & 0.98 & 0.87 & - & - & - & - \\
---- OpenMP Scheduling efficiency (TALP only) & - & - & - & - & - & - & 0.99 & 0.96 \\
---- OpenMP Serialization efficiency (TALP only) & - & - & - & - & - & - & 0.94 & 0.68
\end{tabular}
\caption{Strong scaling-efficiency tables obtained by the different tools.}
\end{sidewaystable}
\FloatBarrier


    \bibliographystyle{spmpsci}
    \bibliography{talp_pages.bib}

\begin{thebibliography}{10}
\providecommand{\url}[1]{{#1}}
\providecommand{\urlprefix}{URL }
\expandafter\ifx\csname urlstyle\endcsname\relax
  \providecommand{\doi}[1]{DOI~\discretionary{}{}{}#1}\else
  \providecommand{\doi}{DOI~\discretionary{}{}{}\begingroup \urlstyle{rm}\Url}\fi

\bibitem{noauthor_extrae_nodate}
Extrae {\textbar} {BSC}-{Tools}

\bibitem{alt_continuous_2024}
Alt, C., Lanser, M., Plewinski, J., Janki, A., Klawonn, A., Köstler, H., Selzer, M., Rüde, U.: A {Continuous} {Benchmarking} {Infrastructure} for {High}-{Performance} {Computing} {Applications}.
\newblock International Journal of Parallel, Emergent and Distributed Systems \textbf{39}(4), 501--523 (2024).
\newblock \doi{10.1080/17445760.2024.2360190}.
\newblock \urlprefix\url{http://arxiv.org/abs/2403.01579}.
\newblock ArXiv:2403.01579 [cs]

\bibitem{anzt_towards_2019}
Anzt, H., Chen, Y.C., Cojean, T., Dongarra, J., Flegar, G., Nayak, P., Quintana-Ortí, E.S., Tsai, Y.M., Wang, W.: Towards {Continuous} {Benchmarking}: {An} {Automated} {Performance} {Evaluation} {Framework} for {High} {Performance} {Software}.
\newblock In: Proceedings of the {Platform} for {Advanced} {Scientific} {Computing} {Conference}, pp. 1--11. ACM, Zurich Switzerland (2019).
\newblock \doi{10.1145/3324989.3325719}.
\newblock \urlprefix\url{https://dl.acm.org/doi/10.1145/3324989.3325719}

\bibitem{bryant_jacamar_nodate}
Bryant, P.: Jacamar {CI} · {GitLab}.
\newblock \urlprefix\url{http://web.archive.org/web/20241129185631/https://gitlab.com/ecp-ci/jacamar-ci}

\bibitem{calotoiu_using_2013}
Calotoiu, A., Hoefler, T., Poke, M., Wolf, F.: Using automated performance modeling to find scalability bugs in complex codes.
\newblock In: Proceedings of the {International} {Conference} on {High} {Performance} {Computing}, {Networking}, {Storage} and {Analysis}, pp. 1--12. ACM, Denver Colorado (2013).
\newblock \doi{10.1145/2503210.2503277}.
\newblock \urlprefix\url{https://dl.acm.org/doi/10.1145/2503210.2503277}

\bibitem{dosimont_monitoring_2024}
Dosimont, D., Houzeaux, G.: Monitoring the development of {CFD} applications on unstable {HPC} platforms (2024).
\newblock \urlprefix\url{http://arxiv.org/abs/2401.08447}.
\newblock ArXiv:2401.08447 [cs]

\bibitem{gamblin_overcoming_2022}
Gamblin, T., Katz, D.S.: Overcoming {Challenges} to {Continuous} {Integration} in {HPC}.
\newblock Computing in Science \& Engineering \textbf{24}(6), 54--59 (2022).
\newblock \doi{10.1109/MCSE.2023.3263458}.
\newblock \urlprefix\url{https://ieeexplore.ieee.org/document/10144946/?arnumber=10144946}.
\newblock Conference Name: Computing in Science \& Engineering

\bibitem{geimer_scalasca_2010}
Geimer, M., Wolf, F., Wylie, B.J.N., Ábrahám, E., Becker, D., Mohr, B.: The {Scalasca} performance toolset architecture.
\newblock Concurrency and Computation: Practice and Experience \textbf{22}(6), 702--719 (2010).
\newblock \doi{10.1002/cpe.1556}.
\newblock \urlprefix\url{https://onlinelibrary.wiley.com/doi/abs/10.1002/cpe.1556}.
\newblock \_eprint: https://onlinelibrary.wiley.com/doi/pdf/10.1002/cpe.1556

\bibitem{brunst_score-p_2012}
Knüpfer, A., Rössel, C., Mey, D.A., Biersdorff, S., Diethelm, K., Eschweiler, D., Geimer, M., Gerndt, M., Lorenz, D., Malony, A., Nagel, W.E., Oleynik, Y., Philippen, P., Saviankou, P., Schmidl, D., Shende, S., Tschüter, R., Wagner, M., Wesarg, B., Wolf, F.: Score-{P}: {A} {Joint} {Performance} {Measurement} {Run}-{Time} {Infrastructure} for {Periscope}, {Scalasca}, {TAU}, and {Vampir}.
\newblock In: H.~Brunst, M.S. Müller, W.E. Nagel, M.M. Resch (eds.) Tools for {High} {Performance} {Computing} 2011, pp. 79--91. Springer Berlin Heidelberg, Berlin, Heidelberg (2012).
\newblock \doi{10.1007/978-3-642-31476-6_7}.
\newblock \urlprefix\url{http://link.springer.com/10.1007/978-3-642-31476-6_7}

\bibitem{lopez_talp_2021}
Lopez, V., Ramirez~Miranda, G., Garcia-Gasulla, M.: {TALP}: {A} {Lightweight} {Tool} to {Unveil} {Parallel} {Efficiency} of {Large}-scale {Executions}.
\newblock In: Proceedings of the 2021 on {Performance} {EngineeRing}, {Modelling}, {Analysis}, and {VisualizatiOn} {STrategy}, pp. 3--10. ACM, Virtual Event Sweden (2021).
\newblock \doi{10.1145/3452412.3462753}.
\newblock \urlprefix\url{https://dl.acm.org/doi/10.1145/3452412.3462753}

\bibitem{martineau_assessing_2017}
Martineau, M., McIntosh-Smith, S., Gaudin, W.: Assessing the performance portability of modern parallel programming models using {TeaLeaf}.
\newblock Concurrency and Computation: Practice and Experience \textbf{29}(15), e4117 (2017).
\newblock \doi{10.1002/cpe.4117}.
\newblock \urlprefix\url{https://onlinelibrary.wiley.com/doi/abs/10.1002/cpe.4117}.
\newblock \_eprint: https://onlinelibrary.wiley.com/doi/pdf/10.1002/cpe.4117

\bibitem{michels_gene-x_2021}
Michels, D., Stegmeir, A., Ulbl, P., Jarema, D., Jenko, F.: {GENE}-{X}: {A} full-f gyrokinetic turbulence code based on the flux-coordinate independent approach.
\newblock Computer Physics Communications \textbf{264}, 107986 (2021).
\newblock \doi{10.1016/j.cpc.2021.107986}.
\newblock \urlprefix\url{https://linkinghub.elsevier.com/retrieve/pii/S0010465521000989}

\bibitem{pillet_paraver_nodate}
Pillet, V., Labarta, J., Cortes, T., Girona, S.: Paraver: {A} tool to visualize and analyze parallel code

\bibitem{saviankou_cubegui_2023}
Saviankou, P., Visser, A., community, C.d.: {CubeGUI}: {Graphical} explorer (2023).
\newblock \doi{10.5281/zenodo.8345207}.
\newblock \urlprefix\url{https://zenodo.org/records/8345207}

\bibitem{schwitanski_--fly_2022}
Schwitanski, S., Tomski, F., Protze, J., Terboven, C., Müller, M.S.: An {On}-the-{Fly} {Method} to {Exchange} {Vector} {Clocks} in {Distributed}-{Memory} {Programs}.
\newblock In: 2022 {IEEE} {International} {Parallel} and {Distributed} {Processing} {Symposium} {Workshops} ({IPDPSW}), pp. 530--540 (2022).
\newblock \doi{10.1109/IPDPSW55747.2022.00093}.
\newblock \urlprefix\url{https://ieeexplore.ieee.org/document/9835251/?arnumber=9835251}

\bibitem{treibig_likwid_2010}
Treibig, J., Hager, G., Wellein, G.: {LIKWID}: {A} {Lightweight} {Performance}-{Oriented} {Tool} {Suite} for x86 {Multicore} {Environments}.
\newblock In: 2010 39th {International} {Conference} on {Parallel} {Processing} {Workshops}, pp. 207--216 (2010).
\newblock \doi{10.1109/ICPPW.2010.38}.
\newblock \urlprefix\url{https://ieeexplore.ieee.org/document/5599200/?arnumber=5599200}.
\newblock ISSN: 2332-5690

\bibitem{wagner_structured_2018}
Wagner, M., Mohr, S., Gimenez, J., Labarta, J.: A {Structured} {Approach} to {Performance} {Analysis} (2018)

\end{thebibliography}
\end{document}